\documentclass[12pt]{article}%
\usepackage{amssymb}
\usepackage{amsmath}
\usepackage{amsfonts}
\usepackage{graphicx}%
\providecommand{\U}[1]{\protect\rule{.1in}{.1in}}

\begin{document}
\bigskip%
\begin{titlepage}
\vspace{.3cm} \vspace{1cm}
\begin{center}
\baselineskip=16pt \centerline{\bf{\Large{ Nonsingular Black Hole}}}%
\vspace{2truecm}
\centerline{\large\bf Ali H.
Chamseddine$^{1,2}$\ , Viatcheslav Mukhanov$^{3,4,5}$\ \ } \vspace{.5truecm}
\emph{\centerline{$^{1}%
$Physics Department, American University of Beirut, Lebanon}}
\emph{\centerline{$^{2}$I.H.E.S. F-91440 Bures-sur-Yvette, France}}
\emph{\centerline{$^{3}%
$Niels Bohr International Academy, Niels Bohr Institute, Blegdamsvej 17, DK-2100 Copenhagen, Denmark}%
}
\emph{\centerline{$^{4}%
$Theoretical Physics, Ludwig Maxmillians University,Theresienstr. 37, 80333 Munich, Germany }%
}
\emph{\centerline{$^{5}%
$MPI for Physics, Foehringer Ring, 6, 80850, Munich, Germany}}
\end{center}
\vspace{2cm}
\begin{center}
{\bf Abstract}
\end{center}
We consider the Schwarzschild black hole and show how,  in a  theory
with limiting curvature,  the physical singularity "inside it" is removed. The resulting spacetime is geodesically
complete. The internal structure of this nonsingular black hole is analogus to Russian nesting dolls.
Namely, after falling into the black hole of radius $r_{g}%
$, an observer, instead of being destroyed at the singularity,
gets for a short time into the region with limiting curvature. After that he re-emerges in the near horizon region
of a  spacetime described by the Schwarzschild metric of a gravitational radius proportional to $r_{g}%
^{1/3}$. In the
next cycle, after passing the limiting curvature, the observer finds himself
within a black hole of even smaller radius proportional to $r_{g}%
^{1/9}$, and so on. Finally after few cycles he will
end up in the spacetime where he remains forever at limiting curvature.
\end{titlepage}%

\section{Introduction}

The problem of singularity within black holes remains, since a long time, as
one of the most intriguing problems in theoretical physics. Although such
singularity is hidden by the event horizon, one can imagine that an observer
can decide (at least in a gedanken experiment) to travel inside the black hole
and the legitimate physical question which arises is: what will this observer
see inside the black hole and in particular as he approaches the singularity?
In case when the black hole has a huge mass he will have more than enough time
to make the needed experiments to measure how the tidal forces are changing.
If General Relativity is valid up to arbitrary high curvatures then the theory
predicts that, irrespective of what any observer will do, he will finally be
destroyed by the infinite curvatures. In fact, assuming universal
applicability of Einstein's theory, and imposing energy dominance conditions
on the state of matter, Hawking and Penrose have proved that space-times with
black holes cannot be geodesically complete $\cite{PenHawk}$. It is well known
that these conditions are not always valid and for instance the condensate of
a scalar field or cosmological constant violate some of them. In this case the
singularity can, in principle, be avoided and the spacetime can become
geodesically complete. For example paper $\cite{FMM}$ considered the
possibility of removing the singularity by forcing the contracting space
inside the black hole to get to the de Sitter bouncing state. This opens the
fascinating possibility of \textquotedblleft gedanken
travelling\textquotedblright\ to another universe via a black hole (of course
only for those who could survive extremely high curvatures at which the bounce
is supposed to take place). However, although this idea by itself does not
contradict any basic physical principles the authors of $\cite{FMM}$ were not
able to provide any concrete example where such an idea could be realized constructively.

Normally the majority of researches redirect the question of singularities to
the yet unknown nonperturbative quantum gravity (which in turn could well be
part of some yet unknown fundamental unified theory). In fact it is clear that
quantum corrections to General Relativity become extremely important at
Planckian curvatures and could easily modify or resolve the singularities.
Therefore, one cannot say that such hopes are completely unjustified. However,
until now, the perturbative treatment of these corrections has led to an
extremely messy picture and did not give, even the slightest constructive
hints of how the problem could be treated and solved in a fully
nonperturbative quantum gravity. Numerous attempts to address this question
did not lead to any reliable progress. Therefore in this paper we will use a
completely different approach. Instead of exploiting quantum effects we will
try to resolve the problem of singularities fully at the classical level by
incorporating the idea of a limiting curvature $\cite{Markov}$, \cite{MukBran}%
, \cite{Mukbransol}, assuming that Einstein's equations are modified at
curvatures well below the Planckian curvature. There is nothing that forbids
this idea because Einstein's equations have been checked experimentally only
for curvatures well below the Planckian ones. If the limiting curvature is
below the Planck value the inevitable quantum effects, due to, for instance,
particle production and vacuum polarization, can be ignored and the theory
will be under control and would remain completely reliable up to the highest
possible curvatures. In particular, in $\cite{BH}$ we have suggested a
concrete theory with limiting curvature and have shown that cosmological
singularities in this theory are fully removed. In this paper we consider how
a black hole is modified in our theory and what happens close to the
singularity inside a black hole. We would like to point out that removing
singularities can have severe consequences for questions broadly discussed in
the literature such as, the so called \textquotedblleft information
paradox\textquotedblright\ and for the fate of remnants of the minimal mass
which can, in principle, survive after the Hawking evaporation is over. We
will discuss these questions in more detail after obtaining the solution for a
nonsingular black hole.

\section{Theory with Limiting curvature}

Consider the theory described by the action \cite{mimetic}, \cite{mimcos}%
\begin{equation}
S=\int\left(  -\frac{1}{2}R+\lambda\left(  g^{\mu\nu}\partial_{\mu}%
\phi\partial_{\nu}\phi-1\right)  +f\left(  \chi\right)  \right)  \sqrt
{-g}d^{4}x,\label{1}%
\end{equation}
where $\chi=\square\phi$, $\lambda$ is a Lagrange multiplier and we have set
$8\pi G=1$. As we have shown in the previous paper $\cite{BH}$ the usual
matter does not play any significant role in resolving anisotropic
singularities. Therefore to simplify the formulae we will omit here its
contribution to $\left(  \ref{1}\right)  .$ It immediately follows from
variation of the Lagrange multiplier $\lambda$ that the scalar $\phi$ always
satisfies the constraint%
\begin{equation}
g^{\mu\nu}\partial_{\mu}\phi\partial_{\nu}\phi=1.\label{2}%
\end{equation}
Therefore the term $f\left(  \chi\right)  ,$ irrespective of any power of
$\chi,$ does not lead to the propagation of extra degrees of freedom which,
otherwise, could be ghosts. The constraint $\left(  \ref{2}\right)  $ imposes
a very strong restriction on the variable $\phi$ and in the synchronous
coordinate system with metric%
\begin{equation}
ds^{2}=dt^{2}-\gamma_{ik}\left(  t,x^{l}\right)  dx^{i}dx^{k},\label{2a}%
\end{equation}
has the most general solution \cite{Landau}
\begin{equation}
\phi=\pm t+A,\label{2b}%
\end{equation}
unless this particular coordinate system does suffer from coordinate
singularities. Thus the field $\phi$ plays the role of time and the constant
of integration $A$ reflects time shift symmetry. In this coordinate system%
\begin{equation}
\chi=\square\phi=\frac{1}{\sqrt{-g}}\frac{\partial}{\partial x^{\mu}}\left(
\sqrt{-g}g^{\mu\nu}\frac{\partial\phi}{\partial x^{\nu}}\right)  =\frac
{\dot{\gamma}}{2\gamma},\label{2c}%
\end{equation}
with $\gamma=\det\gamma_{ik}$ and where by dot we denote time derivative.
Thus, the function $f\left(  \chi\right)  $ allows to introduce, in completely
covariant way, the metric and its first derivative in the \textquotedblleft
game\textquotedblright\ when we try to find simple modification of General
Relativity where singularities can be avoided. In this sense action $\left(
\ref{1}\right)  $ must be treated as a modification of Einstein's gravity. The
only extra new degree of freedom which appears here is mimetic Dark Matter
$\cite{mimetic}$ because constraint $\left(  \ref{2}\right)  $ forces the
longitudinal gravitational field to become dynamical even in the absence of
the usual matter.

We can choose the function $f\left(  \chi\right)  $ in such a way as to bound
the derivative of the metric determinant in the synchronous coordinate system.
Because these derivatives enter in an essential way in the coordinate
independent curvature invariants (see below) this opens the possibility to
have nonsingular modification of gravity. After many trials, the simplest way
we were able to find to construct such a theory is to use a Born-Infeld type
function, where
\begin{equation}
f\left(  \chi\right)  =1-\sqrt{1-\chi^{2}}+g\left(  \chi\right)  , \label{2d}%
\end{equation}
and $\chi$ is restricted by $\chi^{2}\leq1$ for obvious reasons. The function
$g\left(  \chi\right)  $ is less restrictive but it has at least to satisfy
two necessary conditions. First, it must be chosen in such a way as to remove
the $\chi^{2}$ terms in the Taylor expansion of $f\left(  \chi\right)  $
because these would lead to unwanted modification to Einstein's theory at low
curvatures. Second, the function $g\left(  \chi\right)  $ must remove the
singularity in $df/d\chi$ at $\chi=1,$ otherwise the curvature invariants
would blow up at this point. In the theory with $f\left(  \chi\right)  $ given
in $\left(  \ref{2d}\right)  $ the limiting curvature would be of the order of
the Planck curvature, where the quantum effects are extremely important. To
avoid this problem we will assume that the limiting curvature is at least few
orders of magnitude below the Planckian value and this would allow justifying
why vacuum polarization effects and particle production effects could be
ignored. Taking for $g\left(  \chi\right)  $ a function which leads to
particularly simple equations
\begin{equation}
g\left(  \chi\right)  =\frac{1}{2}\chi^{2}-\chi\arcsin\chi
\end{equation}
and introducing limiting curvature, characterized by $\chi_{m}^{2},$ as an
extra free scale in the theory we will take, after rescaling $\chi
\rightarrow\sqrt{\frac{2}{3}}\frac{\chi}{\chi_{m}}$ and $f\rightarrow\chi
_{m}^{2}f$,
\begin{equation}
f\left(  \chi\right)  =\chi_{m}^{2}\left(  1+\frac{1}{3}\frac{\chi^{2}}%
{\chi_{m}^{2}}-\sqrt{\frac{2}{3}}\frac{\chi}{\chi_{m}}\arcsin\left(
\sqrt{\frac{2}{3}}\frac{\chi}{\chi_{m}}\right)  -\sqrt{1-\frac{2}{3}\frac
{\chi^{2}}{\chi_{m}^{2}}}\right)  . \label{3a}%
\end{equation}
As we have already seen in $\cite{BH}$ this choice of $f$ removes
singularities in Friedmann and Kasner universes. In this paper we will
consider what happens with singularities for black holes.

Variation of the action $\left(  \ref{1}\right)  $ with respect to the metric
$g^{\mu\nu}$ gives the modified Einstein's equations%
\begin{equation}
G_{\mu\nu}=R_{\mu\nu}-\frac{1}{2}g_{\mu\nu}R=\tilde{T}_{\mu\nu},\label{3}%
\end{equation}
where
\begin{equation}
\tilde{T}_{\mu\nu}=2\lambda\partial_{\mu}\phi\partial_{\nu}\phi+g_{\mu\nu
}\left(  \chi f^{\prime}-f+g^{\rho\sigma}\partial_{\rho}f^{\prime}%
\partial_{\sigma}\phi\right)  -\left(  \partial_{\mu}f^{\prime}\partial_{\nu
}\phi+\partial_{\nu}f^{\prime}\partial_{\mu}\phi\right)  ,\label{4}%
\end{equation}
characterizes the modification to General Relativity and we have denoted
$f^{\prime}=df/d\chi$. For metric $\left(  \ref{2a}\right)  $ the time-time
and space-space components of the curvature are \cite{Landau}%
\begin{equation}
R_{0}^{0}=-\frac{1}{2}\dot{\varkappa}-\frac{1}{4}\varkappa_{i}^{k}%
\varkappa_{k}^{i},\qquad R_{k}^{i}=-\frac{1}{2\sqrt{\gamma}}\frac{d\left(
\sqrt{\gamma}\varkappa_{k}^{i}\right)  }{dt}-P_{k}^{i},\label{8}%
\end{equation}
where $\varkappa_{k}^{i}=\gamma^{im}\dot{\gamma}_{mk}$, $\varkappa
=\varkappa_{i}^{i}=\dot{\gamma}/\gamma$ and $P_{k}^{i}$ is the three
dimensional Ricci tensor for the metric $\gamma_{ik}$. The corresponding
components of $\tilde{T}_{\nu}^{\mu}$ for solution $\left(  \ref{2b}\right)  $
are
\begin{align}
\tilde{T}_{0}^{0} &  =2\lambda+\chi f^{\prime}-f-\dot{\chi}f^{\prime\prime
},\nonumber\\
\tilde{T}_{k}^{i} &  =\left(  \chi f^{\prime}-f+\dot{\chi}f^{\prime\prime
}\right)  \delta_{k}^{i}.\label{9}%
\end{align}
The $0-0$ equation
\begin{equation}
R_{0}^{0}-\frac{1}{2}R=\tilde{T}_{0}^{0}%
\end{equation}
then takes the form%
\begin{equation}
\frac{1}{8}\left(  \varkappa^{2}-\varkappa_{i}^{k}\varkappa_{k}^{i}+4P\right)
=2\lambda+\chi f^{\prime}-f-\dot{\chi}f^{\prime\prime},\label{10}%
\end{equation}
and the space-space equation
\begin{equation}
R_{k}^{i}=\tilde{T}_{k}^{i}-\frac{1}{2}\tilde{T}_{\alpha}^{\alpha}\delta
_{k}^{i}%
\end{equation}
becomes%
\begin{equation}
\frac{1}{2\sqrt{\gamma}}\frac{\partial\left(  \sqrt{\gamma}\varkappa_{k}%
^{i}\right)  }{\partial t}+P_{k}^{i}=\left(  \lambda+\chi f^{\prime}-f\right)
\delta_{k}^{i}\label{12}%
\end{equation}
Variation of the action with respect to $\phi$ gives
\begin{equation}
\frac{1}{\sqrt{\gamma}}\partial_{0}\left(  2\sqrt{\gamma}\lambda\right)
=\square f^{\prime}=\frac{1}{\sqrt{\gamma}}\partial_{0}\left(  \sqrt{\gamma
}f^{\prime\prime}\dot{\chi}\right)  -\Delta f^{\prime},\label{14}%
\end{equation}
where $\Delta f^{\prime}$ is the covariant Laplacian of $f^{\prime}$ for the
metric $\gamma_{ik}$ and this equation can be used to determine the Lagrange
multiplier $\lambda.$ Up to this point we did not make any assumptions about
the metric $\gamma_{ik}.$ However, for our purposes, it will be enough to
consider only the case when the determinant of the metric is factorizable,
that is, $\gamma\left(  t,x^{i}\right)  =\gamma_{1}\left(  t\right)
\gamma_{2}\left(  x^{i}\right)  .$ Then, both $\chi$ and $\varkappa$ depend
only on time and $\Delta f^{\prime}$ vanishes; hence integrating (\ref{14}) we
obtain
\begin{equation}
\lambda=\frac{C}{2\sqrt{\gamma}}+\frac{1}{2}f^{\prime\prime}\dot{\chi
}.\label{15}%
\end{equation}
where $C$ is a constant of integration corresponding to mimetic cold matter.
Because this matter behaves exactly like dust we can neglect it for the
reasons explained above and set $C=0.$ By subtracting from equation $\left(
\ref{12}\right)  $ one third of its trace we find%
\begin{equation}
\frac{\partial}{\partial t}\left(  \sqrt{\gamma}\left(  \varkappa_{k}%
^{i}-\frac{1}{3}\varkappa\delta_{k}^{i}\right)  \right)  =-2\left(  P_{k}%
^{i}-\frac{1}{3}P\delta_{k}^{i}\right)  \sqrt{\gamma},\label{16}%
\end{equation}
from which it follows that
\begin{equation}
\varkappa_{k}^{i}=\frac{1}{3}\varkappa\delta_{k}^{i}+\frac{\lambda_{k}^{i}%
}{\sqrt{\gamma}},\label{17}%
\end{equation}
where%
\begin{equation}
\lambda_{k}^{i}=-2\int\left(  P_{k}^{i}-\frac{1}{3}P\delta_{k}^{i}\right)
\sqrt{\gamma}dt.\label{17a}%
\end{equation}
and is traceless $\lambda_{i}^{i}=0.$ Substituting expression $\left(
\ref{17}\right)  $ together with $\left(  \ref{15}\right)  $ into $\left(
\ref{10}\right)  $ we obtain
\begin{equation}
\frac{1}{12}\varkappa^{2}+f-\chi f^{\prime}=\frac{\lambda_{k}^{i}\lambda
_{i}^{k}}{8\gamma}-\frac{1}{2}P.\label{18}%
\end{equation}
Taking into account that $\chi=\dot{\gamma}/2\gamma=\varkappa/2$ we infer that
$\left(  \ref{18}\right)  $ is a first order non-linear differential equation
for $\gamma,$ which involves the separate components of the metric only via
the spatial scalar curvature $P$. \ Substituting the function $f$ from
$\left(  \ref{3a}\right)  $ into this equation leads to the particularly
simple equation
\begin{equation}
\chi_{m}^{2}\left(  1-\sqrt{1-\frac{2}{3}\frac{\chi^{2}}{\chi_{m}^{2}}%
}\right)  =\varepsilon,\label{19b}%
\end{equation}
where
\begin{equation}
\varepsilon=\frac{\lambda_{k}^{i}\lambda_{i}^{k}}{8\gamma}-\frac{1}%
{2}P,\label{20a}%
\end{equation}
does not depend on the time derivative of the metric. By squaring $\left(
\ref{19b}\right)  $ and recalling that $\chi=\dot{\gamma}/2\gamma$ \ we
finally arrive at the master equation%
\begin{equation}
\frac{1}{12}\left(  \frac{\dot{\gamma}}{\gamma}\right)  ^{2}=\varepsilon
\left(  1-\frac{\varepsilon}{\varepsilon_{m}}\right)  ,\label{21a}%
\end{equation}
which will be used to analyze the black hole solution and where we have
denoted $\varepsilon_{m}=2\chi_{m}^{2}$.

\section{Schwarzschild solution in General Relativity and the boundary
conditions for $\phi$}

In the empty spherically symmetrical space, solution of Einstein's equations
is unique and is given by the Schwarzschild metric%
\begin{equation}
ds^{2}=\left(  1-\frac{r_{g}}{r}\right)  dt_{S}^{2}-\frac{dr^{2}}{\left(
1-\frac{r_{g}}{r}\right)  }-r^{2}d\Omega^{2}, \label{37}%
\end{equation}
where $r_{g}$ is the gravitational radius and $d\Omega^{2}=d\theta^{2}%
+\sin^{2}\theta d\varphi^{2}$ is the line element on the surface of unit
sphere. This metric is regular both outside the black hole $r>r_{g}$ and
inside the black hole for $r_{g}>r>0$ and only becomes singular on the horizon
at $r=r_{g}.$ Since the singularity occurs \textquotedblleft inside the black
hole\textquotedblright\ it is enough for us to consider only the internal part
of this black hole, where the metric $\left(  \ref{37}\right)  $ is well
applicable and happens to be most convenient for analyzing the internal
structure of a nonsingular black hole. For $r<r_{g}$ the coordinates $r$ and
$t$ exchange their roles and $r$ becomes time-like coordinate while $t_{S}$
becomes space-like one. Inside the black hole the decrease of the
\textquotedblleft radial coordinate\textquotedblright\ from $r=r_{g}$ to $r=0$
corresponds to time increase. Inversely, if we assume that time grows with $r$
then the same Schwarzschild solution describes the white hole, which is just a
time reversed black hole. Let us rename the coordinate in $\left(
\ref{37}\right)  $ as $r\rightarrow r_{g}\tau^{2}$ and $t_{S}\rightarrow R.$
Then inside the Schwarzschild black hole the metric $\left(  \ref{37}\right)
~$becomes
\begin{equation}
ds^{2}=4r_{g}^{2}\tau^{2}N^{-2}\left(  \tau\right)  d\tau^{2}-N^{2}\left(
\tau\right)  dR^{2}-\tau^{4}r_{g}^{2}d\Omega^{2}, \label{38}%
\end{equation}
where%
\begin{equation}
N^{2}\left(  \tau\right)  =\frac{1-\tau^{2}}{\tau^{2}}, \label{39}%
\end{equation}
and where for negative $\tau,$ changing to the interval, $-1\leq\tau\leq0$
describes the collapse \textquotedblleft inside\textquotedblright\ the black
hole until the \textit{spacelike singularity} is reached \textit{at the moment
of time} $\tau=0.$ In fact, the spacetime described by metric $\left(
\ref{38}\right)  $, is obviously non-static and the Riemann squared tensor
equals to
\begin{equation}
R_{\alpha\beta\gamma\delta}R^{\alpha\beta\gamma\delta}=\allowbreak\frac
{12}{\left(  r_{g}\tau^{3}\right)  ^{4}}=\frac{12r_{g}^{2}}{r^{6}},
\label{39a}%
\end{equation}
blows up at the moment of time $\tau=0$ or, as sometimes incorrectly stated,
in the center of the black hole at $r=0$. The Planck curvature is reached at
the moment $\left\vert \tau\right\vert \simeq r_{g}^{-1/3}$ or at $r\simeq
r_{g}^{1/3}.$ Introducing the proper time
\begin{equation}
t=\int2r_{g}\tau N^{-1}\left(  \tau\right)  d\tau=\int\frac{2r_{g}\tau^{2}%
}{\sqrt{1-\tau^{2}}}d\tau=r_{g}\left(  \arcsin\tau-\tau\sqrt{1-\tau^{2}%
}\right)  , \label{40}%
\end{equation}
we can rewrite the metric $\left(  \ref{38}\right)  $ in the form
\begin{equation}
ds^{2}=dt^{2}-a^{2}\left(  t\right)  dR^{2}-b^{2}\left(  t\right)  d\Omega
^{2}, \label{41}%
\end{equation}
where for the Schwarzschild black hole%
\begin{equation}
a^{2}\left(  t\right)  =\frac{1-\tau^{2}\left(  t\right)  }{\tau^{2}\left(
t\right)  },\text{ }b^{2}\left(  t\right)  =\tau^{4}\left(  t\right)
r_{g}^{2}. \label{42}%
\end{equation}
The coordinate system $\left(  \ref{41}\right)  $ is obviously synchronous and
happens to be the most convenient to find a nonsingular generalization of the
Schwarzschild solution in the theory with limiting curvature. Therefore we
will use metric $\left(  \ref{41}\right)  $ and determine the functions
$a^{2}\left(  t\right)  $ and $b^{2}\left(  t\right)  $ which will be modified
in the vicinity of the singularity compared to $\left(  \ref{42}\right)  .$

First of all we notice that at $\chi^{2}\ll\chi_{m}^{2}=\varepsilon_{m}/2$ our
theory coincides with General Relativity in the leading order and therefore
the functions given in $\left(  \ref{42}\right)  $ satisfy equation $\left(
\ref{21a}\right)  $ until we start to approach the limiting curvature. To
determine where the Schwarzschild solution must be valid let us assume that
\begin{equation}
\phi=t+A \label{42a}%
\end{equation}
and calculate%
\begin{equation}
\chi=\square\phi=\frac{\dot{\gamma}}{2\gamma}=\frac{1}{2}\frac{d\ln\left(
a^{2}b^{4}\right)  }{dt}=\frac{\sqrt{1-\tau^{2}}}{4r_{g}\tau^{2}}\frac
{d\ln\left(  \left(  1-\tau^{2}\right)  \tau^{6}r_{g}^{4}\right)  }{d\tau
}=\frac{3-4\tau^{2}}{2r_{g}\tau^{3}\sqrt{1-\tau^{2}}}. \label{43}%
\end{equation}
It then follows from here that for $1>\left\vert \tau\right\vert >\left(
\varepsilon_{m}r_{g}^{2}\right)  ^{-1/6}$ we have $\chi^{2}\ll\varepsilon_{m}$
and the Schwarzschild metric is a good approximation of the exact solution in
$\left(  \ref{21a}\right)  .$ However, one immediately notice that in the near
horizon region (for $\tau^{2}\rightarrow1)$ $\chi^{2}$ grows unbounded for the
Schwarzschild solution although the horizon is nothing more than a coordinate
singularity. It seems that the curvature must grow giving rise to a
\textquotedblleft firewall\textquotedblright\ in our theory, thus completely
modifying Schwarzschild solution, even for large black holes. However, this
\textquotedblleft firewall\textquotedblright\ is completely fake and its
appearance is related to taking the wrong solution $\left(  \ref{42a}\right)
$ for $\phi$ which corresponds to unjustified \textquotedblleft
concentration\textquotedblright\ of this field in the near horizon region that
significantly changes the Schwarzschild solution even outside the
Schwarzschild radius. We have noted above that the solution $\left(
\ref{42a}\right)  $ is a generic solution, \textit{but only} if the
synchronous coordinate system has \textit{no coordinate singularities.
}Obviously the coordinate system $\left(  \ref{41}\right)  $ does not satisfy
this requirement because $\gamma=\left(  1-\tau^{2}\right)  \tau^{6}r_{g}^{4}$
vanishes as $\tau^{2}\rightarrow1.$

To find the synchronous coordinate system which is free of fictitious
coordinate singularities we make a coordinate transformation introducing
instead of $t$ and $R,$ the new coordinates $T$ and $\bar{R}$ defined by
\begin{equation}
T=R+\int\frac{\sqrt{1+a^{2}}}{a}dt,\text{\qquad}\bar{R}=R+\int\frac{dt}%
{a\sqrt{1+a^{2}}} \label{43a}%
\end{equation}
Then in the new \textit{synchronous coordinates} the metric $\left(
\ref{41}\right)  $ becomes%
\begin{equation}
ds^{2}=dT^{2}-\left(  1+a^{2}\right)  d\bar{R}^{2}-b^{2}d\Omega^{2},
\label{43b}%
\end{equation}
where $a^{2}$ and $b^{2}$ are now functions which depend on the argument
$T-\bar{R}.$ For the Schwarzschild solution $\left(  \ref{42}\right)  $ this
metric takes the form%
\begin{equation}
ds^{2}=dT^{2}-\tau^{-2}d\bar{R}^{2}-\tau^{4}r_{g}^{2}d\Omega^{2}, \label{43c}%
\end{equation}
where the relation between $\tau$ and $T-\bar{R}$ can be found by substituting
$\left(  \ref{42}\right)  $ in $\left(  \ref{43a}\right)  $ and taking into
account $\left(  \ref{40}\right)  $:%
\begin{equation}
T-\bar{R}=\frac{2}{3}r_{g}\tau^{3}. \label{43d}%
\end{equation}
This metric describes the Schwarzschild solution in the Lemaitre coordinate
system which is synchronous, regular on the horizon and covers both external
and internal parts of the black hole. Therefore, instead of $\left(
\ref{42a}\right)  ,$ the solution for $\phi$ with correct asymptotic behavior
far away from the black hole is given by%
\begin{equation}
\phi=T=R+\int\frac{\sqrt{1+a^{2}}}{a}dt. \label{44a}%
\end{equation}
Although the Lamaitre coordinates cover the whole manifold and have no
coordinate singularities, they are not very convenient for investigating the
internal structure of nonsingular black holes because the metric components
depend on both space and time coordinates in non separable way. This leads to
equations which have a very complicated structure due to the spatial curvature
terms. Therefore we continue to work in the coordinate system $\left(
\ref{41}\right)  $ but taking the correct solution for $\phi.$ It is easy to
see that $\left(  \ref{44a}\right)  $ satisfy the constraint $\left(
\ref{2}\right)  $ for an arbitrary $a\left(  t\right)  $ as it must be.
Calculating $\chi$ for solution $\left(  \ref{44a}\right)  $ in the coordinate
system $\left(  \ref{41}\right)  $ we find that it is not equal to
$\dot{\gamma}/2\gamma$ anymore and is now given by%
\begin{equation}
\chi=\square\phi=\frac{\dot{\gamma}}{2\gamma}\sqrt{1+\frac{1}{a^{2}}}+\frac
{d}{dt}\sqrt{1+\frac{1}{a^{2}}}. \label{44b}%
\end{equation}
In the case of Schwarzschild black hole we obtain%
\begin{equation}
\chi=\frac{3}{2r_{g}\tau^{3}}, \label{44c}%
\end{equation}
and on the horizon we have $\chi^{2}\ll\varepsilon_{m}$ for $r_{g}%
\gg\varepsilon_{m}^{-1/2}$. Thus for large black holes corrections to
Einstein's equations are negligible on the horizon and the fake firewall does
not arise. Only for very small black holes with a minimal mass determined by
the limiting curvature, the Schwarzschild solution will be completely modified
in our theory. The result is not surprising because in this case the limiting
curvature is already reached on the horizon. Notice that in the case of large
black holes, away from the horizon, $a^{2}\propto\tau^{-2}$ and as we will see
it continues to grow after the bounce in a nonsingular black hole. Therefore
the terms with $1/a^{2}$ in $\left(  \ref{44b}\right)  $ can be neglected once
we are far enough from the original horizon and later. This can be seen by
comparing, for instance, $\left(  \ref{44c}\right)  $ with $\left(
\ref{43}\right)  $ which coincide to order $O\left(  \tau^{2}\right)  $ for
$\tau^{2}\ll1.$ Hence, with good accuracy we can set
\begin{equation}
\chi=\frac{\dot{\gamma}}{2\gamma}, \label{44d}%
\end{equation}
and use $\left(  \ref{21a}\right)  $ to investigate the future of a
nonsingular black hole. If this approximation fails, we would need to work
directly with equation $\left(  \ref{18}\right)  $ with $\chi$ given in
$\left(  \ref{44b}\right)  .$ Fortunately, as we will show, the approximation
holds very well and improves with time and therefore, we can avoid extremely
messy calculations which would be needed otherwise.

Finally, to complete this section we would like to give the approximate
explicit leading order expression for the Schwarzschild metric entirely in
terms of time $t,$ in the near horizon and close to singularity regions. As we
will see this metric will be helpful to understand what happens within the
black hole after reaching the limiting curvature and the bounce.

As seen above, the internal part of the singular black hole is described by
metric $\left(  \ref{41}\right)  ,\left(  \ref{42}\right)  $ for $-1<\tau<0.$
According to $\left(  \ref{40}\right)  $ the proper time $t$ runs in the
interval $-\pi/2<t<0.$ Consider the near horizon region corresponding to
$1+\tau\ll1.$ Then as follows from $\left(  \ref{40}\right)  $%
\begin{equation}
1+\tau\simeq\frac{1}{8}\left(  \frac{\pi}{2}+\frac{t}{r_{g}}\right)
^{2}\equiv\frac{1}{8}\left(  \frac{\bar{t}}{r_{g}}\right)  ^{2} \label{44f}%
\end{equation}
and, in this approximation, the metric takes the form%
\begin{equation}
ds^{2}=d\bar{t}^{2}-\frac{1}{4}\left(  \frac{\bar{t}}{r_{g}}\right)
^{2}dR^{2}-r_{g}^{2}d\Omega^{2}, \label{45h}%
\end{equation}
in the near horizon region for $\bar{t}\ll r_{g}.$ Notice that the numerical
coefficient in front of $dR^{2}$ has no physical meaning because it can be
rescaled by $R\rightarrow\alpha R.$ On the other hand, the coefficient in
front of the angular part of the metric cannot be rescaled and determines the
spatial curvature in the near horizon region which gives a contribution of
order $1/r_{g}^{4}$ to the Riemann squared curvature. For large black holes
the curvature on the horizon is rather small. Now we turn to the region close
to the singularity $\left\vert \tau\right\vert \ll1,$ where
\begin{equation}
t\simeq\frac{2}{3}r_{g}\tau^{3} \label{45}%
\end{equation}
and metric $\left(  \ref{41}\right)  ,\left(  \ref{42}\right)  $ becomes%
\begin{equation}
ds^{2}=dt^{2}-\left(  \frac{t}{t_{0}}\right)  ^{-2/3}dR^{2}-\left(  \frac
{t}{t_{0}}\right)  ^{4/3}r_{g}^{2}d\Omega^{2}, \label{46a}%
\end{equation}
where $t_{0}=2r_{g}/3.$ As one can see from $\left(  \ref{44c}\right)  $ the
limiting curvature is reached when at $t\sim r_{g}\tau^{3}\sim-\varepsilon
_{m}^{-1/2}$ so that $\chi^{2}$ becomes of order $\varepsilon_{m}$ and
$R_{\alpha\beta\gamma\delta}^{2}\sim\varepsilon_{m}^{2}$ (see $\left(
\ref{39a}\right)  $). Before that, the Schwarzschild solution is a good
approximation of the exact solution in the theory with limiting curvature.
Considering the asymptotic expressions $\left(  \ref{45h}\right)  $ and
$\left(  \ref{46a}\right)  $ we can view the evolution of the internal part of
the black hole as a change of one Kasner solution $\left(  \ref{45h}\right)  $
with $p_{i}=\left(  1,0,0\right)  $ in the near horizon region to the other
Kasner solution $\left(  \ref{46a}\right)  $ with $p_{i}^{\prime}=2/3-p_{i}$,
close to the singularity region $\cite{BH}$. This change happens around $t$
$\sim O\left(  1\right)  r_{g}$ and is due to the spatial curvature term
which, as we will see shortly, is only important in this region between the
two asymptotics.

\section{Black hole with limiting curvature}

When the limiting curvature is reached, General Relativity is no longer valid,
and the Schwarzschild solution is modified. To find how and what happens when
we approach the limiting curvature and beyond, we have to solve equation
$\left(  \ref{21a}\right)  ,$ which we quote again for convenience of the
reader
\begin{equation}
\frac{1}{12}\left(  \frac{\dot{\gamma}}{\gamma}\right)  ^{2}=\varepsilon
\left(  1-\frac{\varepsilon}{\varepsilon_{m}}\right)  ,\label{47a}%
\end{equation}
where we now have%
\begin{equation}
\varepsilon=\frac{\lambda_{k}^{i}\lambda_{i}^{k}}{8\gamma}-\frac{1}%
{2}P,\label{47}%
\end{equation}
and%
\begin{equation}
\frac{\lambda_{k}^{i}}{\sqrt{\gamma}}=-\frac{2}{\sqrt{\gamma}}\int\left(
P_{k}^{i}-\frac{1}{3}P\delta_{k}^{i}\right)  \sqrt{\gamma}dt.\label{48}%
\end{equation}
One can easily check that as $\varepsilon_{m}\rightarrow\infty$ the
Schwarzschild solution is the exact solution of these equations. The spatial
curvature components for the metric $\left(  \ref{41}\right)  $ are%
\begin{equation}
P_{1}^{1}=0,\text{ }P_{2}^{2}=P_{3}^{3}=\frac{1}{b^{2}},\text{ }P=\frac
{2}{b^{2}},\label{49}%
\end{equation}
and therefore
\begin{equation}
\frac{\lambda_{k}^{i}}{\sqrt{\gamma}}=-\frac{2\tilde{\lambda}_{\left(
i\right)  }\delta_{k}^{i}}{ab^{2}}F\left(  t\right)  ,\text{ \ \ }F\left(
t\right)  =\int adt,\label{50}%
\end{equation}
where
\begin{equation}
\tilde{\lambda}_{\left(  i\right)  }=\left(  -\frac{2}{3},\frac{1}{3},\frac
{1}{3}\right)  .\label{50a}%
\end{equation}
To determine the constant of integration in $F\left(  t\right)  $ consider the
times $t$ satisfying
\[
\left\vert t\right\vert \gg\varepsilon_{m}^{-1/2},
\]
for which the Schwarzschild solution is valid in the leading approximation.
Then using $\left(  \ref{42}\right)  $ for $a$ and $\left(  \ref{40}\right)  $
to express $dt/d\tau,$ we find
\begin{equation}
\int adt=\int a\left(  \tau\right)  \frac{dt}{d\tau}d\tau=r_{g}\tau
^{2}+C\label{51}%
\end{equation}
and the constant of integration $C$ can be found from equation $\left(
\ref{17}\right)  $ for $\varkappa_{1}^{1},$
\begin{equation}
\varkappa_{1}^{1}=\frac{1}{3}\varkappa+\frac{\lambda_{1}^{1}}{\sqrt{\gamma}%
}=\frac{1}{3}\varkappa+\frac{4\left(  r_{g}\tau^{2}+C\right)  }{3ab^{2}%
}.\label{52}%
\end{equation}
In fact, taking into account that%
\begin{equation}
\varkappa_{1}^{1}=\gamma^{11}\dot{\gamma}_{11}=\frac{d\ln a^{2}}%
{dt},\text{\qquad}\varkappa=\frac{d\ln\left(  a^{2}b^{4}\right)  }%
{dt}\label{53a}%
\end{equation}
and replacing $d/dt$ by the derivative with respect to $\tau,$ equation
(\ref{53a}) simplifies to
\begin{equation}
\frac{\sqrt{1-\tau^{2}}}{2r_{g}\tau^{2}}\frac{d\ln\left(  a/b\right)  }{d\tau
}=\frac{\left(  r_{g}\tau^{2}+C\right)  }{ab^{2}}.\label{53}%
\end{equation}
Substituting for $a$ and $b$ from $\left(  \ref{42}\right)  $ and comparing,
we find that
\begin{equation}
C=-\frac{3}{2}r_{g},\label{53b}%
\end{equation}
and hence,
\begin{equation}
\frac{\lambda_{k}^{i}}{\sqrt{\gamma}}=\frac{2\tilde{\lambda}_{\left(
i\right)  }\delta_{k}^{i}r_{g}}{ab^{2}}\left(  \frac{3}{2}-\tau^{2}\right)
.\label{54}%
\end{equation}
This expression, which we derived in the region where Einstein theory is
applicable, can also be used \textquotedblleft deeply inside the black
hole\textquotedblright\ for $\tau^{2}\ll1$ if we neglect the $\tau^{2}$ term
inside the brackets%
\begin{equation}
\frac{\lambda_{k}^{i}}{\sqrt{\gamma}}=\frac{3\tilde{\lambda}_{\left(
i\right)  }\delta_{k}^{i}r_{g}}{ab^{2}}.\label{54a}%
\end{equation}
Substituting this expression in $\left(  \ref{47}\right)  $ and using $\left(
\ref{49}\right)  $ for the spatial curvature term we finally obtain%
\begin{equation}
\varepsilon=\frac{3r_{g}^{2}}{4a^{2}b^{4}}-\frac{1}{b^{2}}.\label{55}%
\end{equation}
It is clear that for
\begin{equation}
a^{2}b^{2}\ll r_{g}^{2},\label{55a}%
\end{equation}
the spatial curvature term can be neglected. For instance, for the
Schwarzschild black hole this condition takes the form
\begin{equation}
\left(  1-\tau^{2}\right)  \tau^{2}\ll1\label{55b}%
\end{equation}
and hence deeply inside the black hole $\left(  \tau^{2}\ll1\right)  $ and
close to the horizon $\left(  \left(  1-\tau^{2}\right)  \ll1\right)  $ the
spatial curvature term in $\left(  \ref{55}\right)  $ is negligible. Thus
ignoring this term and taking into account that $\gamma=a^{2}b^{4}\sin
^{2}\theta=\gamma_{t}\sin^{2}\theta$, hence, $\dot{\gamma}/\gamma=\dot{\gamma
}_{t}/\gamma_{t}$ and after substitution of $\left(  \ref{55}\right)  $ in
$\left(  \ref{47a}\right)  $ we obtain the equation%
\begin{equation}
\left(  \frac{\dot{\gamma}_{t}}{\gamma_{t}}\right)  ^{2}=\frac{9r_{g}^{2}%
}{\gamma_{t}}\left(  1-\frac{3r_{g}^{2}}{4\varepsilon_{m}\gamma_{t}}\right)
,\label{56}%
\end{equation}
which can be easily integrated to give the solution
\begin{equation}
\gamma_{t}=\frac{3r_{g}^{2}}{4\varepsilon_{m}}\left(  1+3\varepsilon_{m}%
t^{2}\right)  .\label{57}%
\end{equation}
The corresponding metric components $\ a^{2}\left(  t\right)  $ and
$b^{2}\left(  t\right)  $ can be obtained directly from $\left(
\ref{17}\right)  .$ For instance the equation for $\varkappa_{1}^{1}$ takes
the following explicit form%
\begin{equation}
\frac{d\ln a^{2}}{dt}=\frac{1}{3}\frac{\dot{\gamma}_{t}}{\gamma_{t}}%
+\frac{2r_{g}}{\sqrt{\gamma_{t}}}.\label{57a}%
\end{equation}
Integrating this equation, using $\gamma_{t}$ from $\left(  \ref{57}\right)
,$ we find
\begin{equation}
a^{2}\left(  t\right)  =\left(  \frac{3r_{g}^{2}}{4\varepsilon_{m}}\left(
1+3\varepsilon_{m}t^{2}\right)  \right)  ^{1/3}\exp\left(  \frac{4}{3}\left(
\sinh^{-1}\left(  \sqrt{3\varepsilon_{m}}t\right)  +\ln\left(  \frac{4}%
{3}\sqrt{3\varepsilon_{m}}\right)  \right)  \right)  ,\label{58}%
\end{equation}
where the constant of integration is fixed by requiring that before the bounce
for $\left\vert t\right\vert \gg\varepsilon_{m}^{-1/2}$ the asymptotic form of
the solution must be given by $\left(  \ref{46a}\right)  .$ Similarly we
obtain
\begin{equation}
b^{2}\left(  t\right)  =\left(  \frac{3r_{g}^{2}}{4\varepsilon_{m}}\left(
1+3\varepsilon_{m}t^{2}\right)  \right)  ^{1/3}\exp\left(  -\frac{2}{3}\left(
\sinh^{-1}\left(  \sqrt{3\varepsilon_{m}}t\right)  +\ln\left(  \frac{4}%
{3}\sqrt{3\varepsilon_{m}}\right)  \right)  \right)  .\label{59}%
\end{equation}
Thus, the singularity is avoided and instead of it we have a bounce of
duration $\Delta t\simeq\varepsilon_{m}^{-1/2}$. During this time the
curvature is not much different from the limiting curvature but drastically
drops after that. In fact, as follows from $\left(  \ref{58}\right)  $ and
$\left(  \ref{59}\right)  $, after the bounce for $t\gg\varepsilon_{m}^{-1/2}$
the metric is
\begin{equation}
ds^{2}=dt^{2}-Q_{0}^{2}\left(  \frac{t}{t_{0}}\right)  ^{2}dR^{2}-\frac
{1}{Q_{0}}r_{g}^{2}d\Omega^{2},\label{61}%
\end{equation}
where $Q_{0}=\left(  \frac{16}{3}\varepsilon_{m}r_{g}^{2}\right)  ^{2/3}.$ If
the size of the black hole $r_{g}$ is much larger than $\varepsilon_{m}%
^{-1/2}$ then $Q_{0}\gg1.$ The asymptotic form $\left(  \ref{61}\right)  $ is
valid only when the spatial curvature could be neglected and the condition
$\left(  \ref{55a}\right)  $ is satisfied. It holds until the time $t\sim
r_{g}/Q_{0}^{1/2}$ where it is violated. For $t\gg\varepsilon_{m}^{-1/2}$ we
have $\chi^{2}\ll\varepsilon_{m}.$ Moreover, using the formulae from the
Appendix it can be readily checked that for the solution $\left(
\ref{61}\right)  $%
\begin{equation}
R_{\alpha\beta\gamma\delta}^{2}\sim\frac{Q_{0}^{2}}{r_{g}^{4}}\sim\left(
\frac{\varepsilon_{m}}{r_{g}}\right)  ^{4/3}\label{62}%
\end{equation}
and it follows that $R_{\alpha\beta\gamma\delta}^{2}\ll\varepsilon_{m}^{2}$
for large black holes with $r_{g}\gg\varepsilon_{m}^{-1/2}$. Hence, at these
times corrections to Einstein equations are negligible and $\left(
\ref{61}\right)  $ must be a solution of Einstein equations in empty space for
a spherically symmetric metric. We know, however, that such solution is unique
and is described by the Schwarzschild metric. In fact, rescaling
$R\rightarrow\tilde{R}=3Q_{0}^{1/2}R$ and introducing
\begin{equation}
R_{g_{1}}=\frac{r_{g}}{Q_{0}^{1/2}}=\frac{r_{g}^{1/3}}{\left(  16\varepsilon
_{m}/3\right)  ^{1/3}},\label{63}%
\end{equation}
we can rewrite $\left(  \ref{61}\right)  $ as
\begin{equation}
ds^{2}=dt^{2}-\frac{1}{4}\left(  \frac{t}{R_{g_{1}}}\right)  ^{2}d\tilde
{R}^{2}-R_{g_{1}}^{2}d\Omega^{2}.\label{64}%
\end{equation}
Comparing this metric to $\left(  \ref{45h}\right)  $ we can identify its
spacetime with the inner side of the near horizon asymptotic of the
Schwarzschild solution with gravitational radius $R_{g_{1}}\propto r_{g}%
^{1/3}.$ As pointed out above, at the moment of time $t\sim R_{g_{1}}$ the
spatial curvature term in $\left(  \ref{55}\right)  $ becomes dominant and
changes the asymptotic solution $\left(  \ref{64}\right)  $ to another one
which can be written by analogy with $\left(  \ref{46a}\right)  .$ We simply
take into account that in the corresponding Schwarzschild black hole with
radius $R_{g_{1}}$ the singularity would be reached at $t=\frac{\pi}%
{2}R_{g_{1}}$ and we can write
\begin{equation}
ds^{2}=dt^{2}-\left(  \frac{t-\frac{\pi}{2}R_{g_{1}}}{t_{1}}\right)
^{-2/3}d\tilde{R}^{2}-\left(  \frac{t-\frac{\pi}{2}R_{g_{1}}}{t_{1}}\right)
^{4/3}R_{g_{1}}^{2}d\Omega^{2},\label{65}%
\end{equation}
where $t_{1}=\frac{2}{3}R_{g_{1}}.$ This solution is valid until the limiting
curvature is reached, that is, for $\frac{\pi}{2}R_{g_{1}}-t\gg\varepsilon
_{m}^{-1/2}.$ After we start to approach the limiting curvature the solution
changes, and it is described by the formulae $\left(  \ref{58}\right)
,\left(  \ref{59}\right)  $ with the obvious replacements $r_{g}\rightarrow
R_{g_{1}},$ $t-\frac{\pi}{2}R_{g_{1}}.$ To return to the original scale factor
$a^{2}\left(  t\right)  $ we rescale $\tilde{R}$ back to $R.$ As a result,
after \textit{the second bounce,} we again re-emerge inside the near horizon
region described by the metric
\begin{equation}
ds^{2}=dt^{2}-9Q_{0}Q_{1}^{2}\left(  \frac{t-\frac{\pi}{2}R_{g_{1}}}{t_{1}%
}\right)  ^{2}dR^{2}-\frac{1}{Q_{1}}R_{g_{1}}^{2}d\Omega^{2},\label{66}%
\end{equation}
for $t-\frac{\pi}{2}R_{g_{1}}\gg\varepsilon_{m}^{-1/2},$ \ where%
\begin{equation}
Q_{1}=\left(  \frac{16}{3}\varepsilon_{m}R_{g_{1}}^{2}\right)  ^{2/3}=\left(
\frac{16}{3}\varepsilon_{m}\frac{r_{g}^{2}}{Q_{0}}\right)  ^{2/3}=Q_{0}%
^{1/3}.\label{68}%
\end{equation}
Obviously, the metric $\left(  \ref{66}\right)  $ describes the near horizon
Schwarzschild geometry with the gravitational radius%
\begin{equation}
R_{g_{2}}=\frac{R_{g_{1}}}{Q_{1}^{1/2}}=\frac{r_{g}}{Q_{0}^{\frac{1}{2}\left(
1+\frac{1}{3}\right)  }}=\frac{r_{g}^{1/9}}{\left(  16\varepsilon
_{m}/3\right)  ^{4/9}}.\label{69}%
\end{equation}
Repeating the steps above we find that the spacetime structure inside the
nonsingular black hole is similar to \textquotedblleft a Russian nesting
doll\textquotedblright. Namely, its geometry is a time sequence of the
internal Schwarzschild geometries separated by \textquotedblleft layers with
limiting curvature\textquotedblright\ of width $\Delta t\simeq\varepsilon
_{m}^{-1/2}.$ The Scharzschild radii characterizing these subsequent
geometries decrease and are proportional to $r_{g},$ $r_{g}^{1/3},$
$r_{g}^{1/9},$ $r_{g}^{1/27},$....etc. After the $n+1$ bounce (the first
bounce takes place at $t=0$), which happens at the moment%
\begin{equation}
T_{n}=\frac{\pi}{2}\left(  R_{g_{1}}+R_{g_{2}}+...+R_{g_{n}}\right)
\label{70}%
\end{equation}
the gravitational radius is equal to
\begin{equation}
R_{g_{\left(  n+1\right)  }}=\frac{R_{g_{\left(  n\right)  }}}{Q_{n}^{1/2}%
}=\frac{r_{g}}{Q_{0}^{\frac{1}{2}\left(  1+\frac{1}{3}+...+\frac{1}{3^{n}%
}\right)  }}=\frac{1}{\left(  16\varepsilon_{m}/3\right)  ^{1/2}}\exp\left(
\frac{1}{4\cdot3^{n}}\ln Q_{0}\right)  \label{71}%
\end{equation}
and when the gravitational radius becomes comparable with the minimal possible
one
\begin{equation}
R_{g_{\min}}=\frac{1}{\left(  16\varepsilon_{m}/3\right)  ^{1/2}}\label{72}%
\end{equation}
the approximations we used to obtain the picture described above breaks down.
In fact, after
\begin{equation}
n_{\max}\sim\ln\ln Q_{0}\label{73}%
\end{equation}
bounces the width of the layers with limiting curvature is of the order of the
size of the black hole and we cannot use anymore the Schwarzschild solution in
between the layers. After that the limiting curvature is reached and never
drops to small values. The corresponding geometry is similar to the one which
describes the minimal black hole in our theory $\cite{CMF}.$

\section{Summary and speculations}

We have shown that in the theory with limiting curvature the internal
structure of a black hole is significantly modified compared to a singular
Schwarzschild black hole. Namely, the curious observer who decides to travel
inside the Schwarzschild eternal black hole after first crossing the horizon
will find himself in a non-static space of infinite volume (for eternal black
hole), but exists for finite time $t\sim r_{g}.$ At the beginning the
curvature of large black holes is very low but grows and finally, after time
$t\sim r_{g}$,\ becomes infinite and one ends up in a singularity, which
happens not \textquotedblleft at the point in the center of black
hole\textquotedblright\ but at the moment of time $t=0.$ In this sense the
evolution and singularity within a black hole is similar to a Kasner universe.
The spacetime in this case is not geodesically complete. In our theory with
limiting curvature, Einstein equations are only significantly modified when
the curvature starts to approach its limiting value. The singularity is
removed and the curvature does not grow indefinitely. In fact, the singularity
is replaced by a \textquotedblleft time layer\textquotedblright\ of duration
$\Delta t\sim\varepsilon_{m}^{-1/2}\,,$ which would be of the order of Planck
time if the limiting curvature would be the Planckian one. After that the
curvature drops down to the value which an observer would find immediately
after crossing the horizon of the smaller black hole of radius $r_{g}^{1/3}$.
The subsequent evolution repeats the previous cycle but this time inside a
black hole of this smaller radius. Once again, instead of ending at the
singularity we pass through a layer of limiting curvature and find ourselves
inside a black hole of even smaller radius $\sim r_{g}^{1/9}$ and so on.
Finally when the size of the black hole becomes of the order of the width of a
time layer $\sim\varepsilon_{m}^{-1/2},$ we end inside the black hole of
minimal possible mass and stay there forever at limiting curvature. Notice
that the number of the \textquotedblleft layers\textquotedblright\ which we
have to pass to reach inside this minimal black hole is not big even for large
black holes. For instance, for a galactic mass black hole of radius $r_{g}%
\sim10^{49}$ (in Planck units) after the crossing of limiting curvature we
find ourselves in  black holes of radii $r_{g}^{1/3}\sim10^{16},$ $r_{g}%
^{1/9}\sim10^{5},$ $r_{g}^{1/27}\sim10^{2}$ correspondingly. Finally at the
fourth layer $r_{g}^{1/81}\sim O\left(  1\right)  $ and we cannot trust
anymore the approximations used to arrive at the above picture and we end up
within a minimal black hole at limiting curvature, which after that never
drops significantly. The spacetime of a nonsingular black hole is geodesically
complete and the singularity problem is resolved.

For an evaporating black hole the derivation of Hawking radiation remains
unchanged for a large black hole \cite{GH}. However, when it reaches the
minimal size of order $\varepsilon_{m}^{-1/2}$ the near horizon geometry
changes and we expect that the minimal remnants of it must be stable. This
question obviously requires further investigation $\cite{CMF}$. If we take the
limiting curvature, which is a free parameter in our theory, to be at least a
few orders of magnitude below the Planck scale, the answer to it can be
obtained using standard methods of quantum field theory in external
gravitational field. In fact, in this case the unknown nonperturbative quantum
gravity does not play an essential role and its need in such a case becomes
unclear because the uncontrollable Plankian curvatures are never reached. This
opens up the possibility of resolving the information paradox without
involving the \textquotedblleft mysteriously imprinted\textquotedblright%
\ correlations in Hawking radiation which is supposed to take care of
\ returning all information back to the Minkowski space after disappearance of
the black hole. In our case the smallest black hole remnant has enough space
\textquotedblleft inside it\textquotedblright\ to hide all the information
about the original matter from which the black hole was formed together with
the information about the negative energy quanta (with respect to an outside
observer) which never escapes from the black hole and reduce its mass in the
process of Hawking evaporation. The evolution in this case remains unitary on
complete Cauchy hypersurfaces which inevitably goes inside the black hole
remnant. The picture here is very similar to the one described as a possible
option in $\cite{FMM}.$ The content of the minimal mass black hole can be
significantly different depending on the way how the remnant was formed.
However, an infinite degeneracy of the black hole remnants is completely
irrelevant for an outside observer who calculates, for instance, the
scattering processes with participation of these minimal black holes, because
this degeneracy is entirely related to events which happen in the absolute
future of this observer.

\section{Appendix}

For convenience of the reader we quote below the explicit expressions for
curvature invariants which can be used to verify statements about the behavior
of the curvature in a nonsingular black hole with the metric $\left(
\ref{41}\right)  $. The scalar curvature is given by the expression%
\[
R=-\dot{\varkappa}-\frac{1}{3}\varkappa^{2}-\frac{2}{3}\frac{F^{2}}{\gamma
}-\frac{2}{b^{2}},
\]
where $F=\int adt$. The square of the Ricci tensor is given by%
\[
R_{\alpha\beta}R^{\alpha\beta}=\frac{1}{3}\dot{\varkappa}^{2}+\frac{1}%
{6}\varkappa^{2}\dot{\varkappa}+\frac{1}{36}\varkappa^{4}+\frac{2}{3}\left(
\dot{\varkappa}+\frac{1}{6}\varkappa^{2}\right)  \frac{F^{2}}{\gamma}+\frac
{4}{9}\frac{F^{4}}{\gamma^{2}}+\frac{1}{b^{2}}\left(  \frac{2}{3}%
\dot{\varkappa}+\frac{1}{3}\varkappa^{2}\right)  +\frac{4a^{2}}{3\gamma}%
\]
and the square of the Riemann tensor is
\begin{align*}
R_{\alpha\beta\gamma\delta}R^{\alpha\beta\gamma\delta}  &  =\left(
\frac{\varkappa^{4}}{54}+\frac{\dot{\varkappa}^{2}}{3}+\frac{\varkappa^{2}%
\dot{\varkappa}}{9}\right)  +\frac{2}{9}\left(  4\dot{\varkappa}+\varkappa
^{2}\right)  \frac{F^{2}}{\gamma}-\frac{16\varkappa}{27}\frac{F^{3}}%
{\gamma^{\frac{3}{2}}}\\
&  +\frac{4F^{4}}{3\gamma^{2}}+\frac{8a^{2}}{3\gamma}+\frac{16a}{9}\frac
{F^{2}}{\gamma^{\frac{3}{2}}}-\frac{8\varkappa aF}{9\gamma}+\frac{1}{b^{2}%
}\left(  \frac{4}{b^{2}}+\frac{2\varkappa^{2}}{9}+\frac{8F^{2}}{9\gamma}%
-\frac{8\varkappa F}{9\sqrt{\gamma}}\allowbreak\right)
\end{align*}

\bigskip

\textbf{{\large {Acknowledgments}}}

The work of A. H. C is supported in part by the National Science Foundation
Grant No. Phys-1518371. The work of V.M. is supported in part by Simons
Foundation grant 403033TRR 33 \ and \textquotedblleft The Dark
Universe\textquotedblright\ and the Cluster of Excellence EXC 153
\textquotedblleft Origin and Structure of the Universe\textquotedblright%
.\bigskip\ 

\bigskip

\bigskip

\bigskip

\begin{thebibliography}{99}                                                                                               %


\bibitem {PenHawk}S. W. Hawking and R. Penrose, \textit{The singularities of
gravitational collapse and cosmology, }Proc. Roy. Soc. Lond. A. \textbf{314}
(1970) 529.

\bibitem {FMM}V. Frolov, M. Markov and V. Mukhanov, \textit{Through a Black
Hole into a New Universe? }Phys. Lett. \textbf{B216 }(1989) 272.

\bibitem {Markov}M. Markov, Pis'ma Zh. Eskp. Teor. Fiz. \textbf{36 }(1982)
214; \textbf{46 }(1987) 341 [JETP Lett. \textbf{36 }(1982) 265; \textbf{46
}(1987) 431.

\bibitem {MukBran}V. Mukhanov and R. Brandenberger, \textit{A Nonsingular
Universe, }Phys. Rev. Lett. \textbf{68 (}1992) 1969.

\bibitem {Mukbransol}V. Mukhanov, R. Brandenberger, and A. Sornborger,
\textit{A Cosmological Theory without Singularities, }Phys. Rev. D \textbf{48
}(1993) 1629.

\bibitem {BH}A. H. Chamseddine, V. Mukhanov, \textit{Resolving Cosmological
Singularities.}

\bibitem {mimetic}A. H. Chamseddine and V. Mukhanov, \textit{Mimetic Dark
Matter, }JHEP \textbf{1311 }(2013) 135.

\bibitem {mimcos}A. H. Chamseddine, V. Mukhanov and A. Vikman,
\textit{Cosmology with Mimetic Matter, }JCAP \textbf{1406 }(2014) 017.

\bibitem {Landau}L. Landau and E. Lifshitz, \textit{The Classical Theory of
Fields} fourth edition, Butterworth, Heinemann, 1980.

\bibitem {GH}G. W. Gibbons and S. W. Hawking, \textit{Cosmolgical event
horizons, thermodynamics and particle creation, }Phys. Rev. D \textbf{15
}(1977) 2738.

\bibitem {CMF}A. H. Chamseddine, V. Mukhanov, \textit{In Preparation.}
\end{thebibliography}
\end{document}